\documentclass{ws-p8-50x6-00}

\begin{document}

\title{Heavy Ion Physics at LHC}

\author{G. VALENTI}

\address{Istituto Nazionale di Fisica Nucleare Sez di Bologna,
Via Berti Pichat 6/22, 40127 Bologna, ITALY\\ 
E-mail: Giovanni.Valenti@cern.ch}


\maketitle

\abstracts{
The study of heavy ion interactions constitutes an important part of the 
experimental program outlined for the Large Hadron
Collider under construction at CERN and expected to be operational by 2006.
ALICE~\cite{al} is the single detector having the capabilities 
to explore at the same time most of the characteristics of high 
energy heavy ion interactions.
Specific studies of jet quenching and quarkonia production, 
essentially related to $\mu$ detection are also 
planned by CMS~\cite{cms}.}

\section{Introduction}

The main goal of ultrarelativistic heavy ion collisions
is to study the thermodynamics of strongly interacting 
plasma of quarks and gluons. 
At  high energy densities, QCD calculations predict a deconfinement 
of quark and gluons resulting in a new
state of matter: the Quark Gluon Plasma (QGP).
Recent lattice calculations~\cite{ka} show that for a system of 
vanishing baryonic
density, the transition temperature between the confined hadron and the 
parton phases is $T_c \sim$ 150 - 200 MeV and the critical energy density 
$\varepsilon_c$ of the order 1 - 3 GeV/$fm^3$. Furthermore the transition 
is expected to be accompanied by the restoration of the chiral symmetry 
of the QCD interactions, spontaneously broken at low
energy density.

Relativistic Heavy Ion Collisions (HIC) offer an hitherto
unprecedented possibility to investigate experimentally
the QGP phase diagram.  In fact, lead ion collisions at the 
LHC ($\sqrt{s}$ = 5.5 A TeV) will allow 
exploration of a new domain in energy density,
kinematic and dynamic conditions, typically an order of magnitude 
higher then what has been observed at SPS
energies at CERN ($\sqrt{s} \sim$ 20 A GeV)~\cite{qm01}.
At the LHC the collision evolution will be governed by hard parton
production due to concomitant increase of p-p total inelastic cross section
and $A^{4/3}$ scaling in the collision of large nuclei with 
a shorter production time of the order of
$\tau_h$ $\simeq$ 0.1 fm/c.  Soft processes (hadronization) will be
completed later, typically at $\tau_s$ $\sim$ 
1/$\Lambda_{QCD}$ $\sim$ 1. fm/c.
Furthermore, as the system created by the initial collision
will be dominated by gluons~\cite{macl} (up to 80\%) characterised
by small $x = 2p_T/\sqrt{s}$, the baryon density in the central rapidity region will be substantially diluted.

\section{The Observables}

The system produced in a relativistic heavy ion collision (Pb-Pb)
is extended and collective with a large volume ($\gg 1000 fm^3$) and a 
long lifetime ($>$ 1 fm/c).  Such a system can only be controlled as 
far as the initial parameters i.e. the ion mass and the collision
energy. A posteriori events can be classified (triggered) according to the 
total multiplicity and energy observed at zero degree (c.m.s.)
this being strongly correlated to the collision impact parameter.
The study of the space-time evolution of the collision must, 
therefore, be inferred from the observables measured in the final state.

The strategy ALICE will adopt is to measure at the same time 
in the same detector global characteristics of the collision like 
multiplicity, forward and transverse energy as well as 
specific final state observables like hadron $p_t$ distributions, 
particle ratios, strangeness production indicative of
the expected phase transition and real and virtual photons, 
charmonium production, vector meson properties (mass)
indicative of the QGP phase during the very first stage of
the collision.

Some of the final state characteristics are controlled by the very 
first instance of the ion collision. This is the case
for the direct photons, electron pairs, J/$\Psi$ and open charm
production.

Direct photons~\cite{phot} originate from the quark-gluon 
phase, it is therefore
expected that their momentum distribution is shaped by the 
thermodynamical state at the very early stage of the collision.

Electron pairs mass distributions show, as seen at the SPS, an
interesting anomalous excess in the low mass region 250 to 700 MeV
not observed in p-nucleus reactions. Suggestions vary from $\rho$
meson mass broadening to modification of its mass due to the 
surrounding dense hadronic medium~\cite{esk}.
J/$\Psi$ suppression~\cite{jpsi} is considered a 
clear signal of the screening
of the binding of c$\bar{c}$ formed by gluon fusion within the 
quark-gluon plasma.

Open charm~\cite{charm} will mostly be produced in 
the early interaction stage 
(because of the heavy mass) and will survive to the final state.

The bulk of the particle in the final state will however derive from 
hadronic production occurring after the freeze out of the
system. The signals are usually very strong to the point that some 
characters of the reactions can be studied on an event by event base
where correlations and non statistical fluctuations are predicted 
for critical phenomena in a phase transition scenario.\\

\section{ALICE}

ALICE is a dedicated heavy ion detector aimed at the study of hadrons, 
leptons and photons produced in Pb-Pb collisions at the LHC.  The detector 
is designed to cope with the highest particle multiplicity anticipated 
for Pb-Pb reactions at $\sqrt{s}$ = 5.5 A TeV: dN$_{ch}/dy~\sim$ 8000.
Lower mass ion collisions like Ca-Ca, as well as p-p and p-nucleus are 
also included in the LHC program as they will provide reference data 
and a way of varying the initial energy density.

The main components of the ALICE detector shown in fig. 1 
consists of a central part embedded
in the 0.2 (0.4) T field of the LEP L3 magnet, a dimuon detector, a 
forward multiplicity detector and a zero degree calorimeter in the 
forward region some 100 m from the interaction region.
The central rapidity ( $\pm$ 0.9) is instrumented with:
\begin{itemize}
\item The inner tracking system ITS, a high resolution silicon tracker 
consisting of 6 cylindrical layers 
(inner radius 4 cm and outer radius 44 cm) whose basic functions
are the determination of the primary and secondary vertices necessary
to study charm and hyperon decays, identification and 
tracking for low momentum particles ($\leq$ 100 MeV/c), improve the 
momentum and angle resolution of the TPC.
\item The Time Projection Chamber (TPC), ALICE main tracking system
with an inner radius of 90 cm, determined by the maximum sustainable hit density,
and outer radius 250 cm determined by the minimum track length
for a 10\% dE/dx measurement.
\item The particle identification complex consisting of a Transition Radiation
Detector (TRD) to identify electrons with momentum greater then 1 GeV/c.
A Time Of Flight (TOF) system consisting of 160K channels of multi gap RPC
located at 3.5 meters radius with an intrinsic time resolution better
then 100 ps. Complemented by two smaller coverage detector, the High Momentum
Particle Identification Detector (HMPID), a proximity focus rich
detector for high $p_t$ particle identification and the Photon
Spectrometer (PHOS) a $PBWO_4$ calorimeter capable to
detect in a about 5\% of the phase space of the central 
barrel $\gamma$'s and high $p_t$ neutral mesons.
\end{itemize}
The forward dimuon spectrometer arm covers $\eta$ = 2.5 - 4. and consists 
of a composite absorber to stop most of the hadrons, photons and electrons.
Muons trajectories bent by a 3 Tm magnetic dipole are then measured by 10 
plane tracking system consisting of cathode pad chambers, and finally
identified by four more detector planes located behind a second iron 
absorber.
The set-up is completed by the forward multiplicity detector (FMD), 
a preshower detector with fine granularity and full azimuthal 
coverage in the pseudorapidity region 1.8 $\leq \eta \leq$ 2.6 and 
the forward electromagnetic and hadronic 
calorimeter, CASTOR, having full azimuthal coverage in the 
pseudorapidity region 5.6 $\leq \eta \leq$ 7.2.  
 
\section{CMS}

The Compact Muon Solenoid detector is constructed around a large magnet
(14 m long and 3 m radius) operated at a magnetic field of 4. T.
The coverage of the central tracker (layers of silicon pixels and strips),
electromagnetic and hadronic calorimeters followed by $\mu$ detector, relevant to
the heavy ion physics run is $\pm$ 2.4~\cite{cmsd}.
The good momentum resolution (p $\geq$ 3.5 GeV/c) and $\mu$ 
identification will be used to study:  

i) Suppression within quark-antiquark resonances ($\Upsilon, \Upsilon', \Upsilon'',
J/\Psi, \Psi'$).

ii) Z $\rightarrow$ \, $\mu^+$ $\mu^-$ production (jet tagging, 
reference for $\Upsilon$ suppression)

iii) Jet production: jet pairs, Z + jet, $\gamma$ + jet, possibly
a study of jet quenching.

\section{Conclusions}

ALICE covers in depth most of the experimental parameters relevant to
the study of heavy ion collision at the LHC. The CMS detector will 
contribute to the study of jet quenching and quarkonia production.

\vspace{1cm}\noindent

\begin{figure}[h]
\begin{center}
\epsfxsize=30pc
\epsfbox{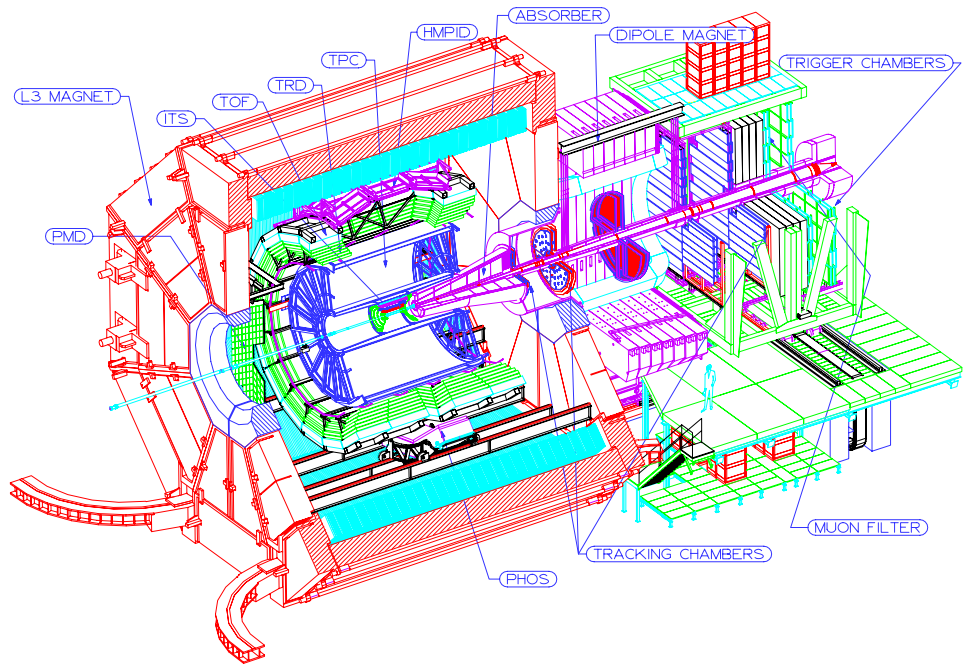}
\caption{Artist view of the ALICE detector.  \label{fig:radish}}
\end{center}
\end{figure}

\end{document}